\documentstyle[prl,aps,twocolumn,epsf]{revtex}
\begin{document}
\title{Current Switch by Coherent Trapping of Electrons in Quantum Dots}
\author{T. Brandes$^1$, F. Renzoni$^2$}
\address{$^1$Universit\"at Hamburg, 1. Institut f\"ur  Theoretische
Physik, Jungiusstr. 9, D-20355 Hamburg, Germany. Present and permanent address: 
Department of Physics, University of Manchester Institute of Science and Technology (UMIST), P.O. Box 88, 
Manchester M60 1QD, UK.\\
$^2$Universit\"at Hamburg, Institut f\"ur Laserphysik, Jungiusstr. 9, D-20355 Hamburg, Germany}
\date{\today}
\maketitle
\draft
\begin{abstract}
We propose a new transport mechanism through tunnel--coupled quantum dots based on
the coherent population trapping effect. Coupling to an excited level by the 
coherent radiation of two microwaves can lead to an extremely narrow current
antiresonance. The effect can be used to determine interdot dephasing rates
and is a mechanism for a very sensitive, optically controlled current switch.
\end{abstract}
The analogy between real and artificial atoms (quantum dots) suggests
the transfer of concepts from atomic physics to ultrasmall semiconductor structures. 
If methods like optical coherent control are combined with the tunability of quantum dots,
basic quantum mechanical effects like preparation in a 
superposition of states and quantum interference can be realized and controlled
in artificial microscopic devices.
The interaction with light has been used to create
coherent superpositions of states in single \cite{Bonetal98} and 
double quantum dots \cite{Blietal98a}. Furthermore, external radiation
fields lead to non--linear electron transport effects like photo--assisted tunneling and photo--sidebands
\cite{Kouetal94,Oosetal98}. 

In this Letter, we propose a new transport mechanism through tunnel--coupled quantum dots based on
the coherent population trapping effect, a well--known effect in
atomic laser spectroscopy \cite{Arimondo96}. We predict that the interaction
with coherent light of two frequencies can be used to pump a current through a double dot.
As a function of the relative detuning of the two frequencies the current shows 
an extremely narrow antiresonance, i.e. an optically controlled 
abrupt transition from a conducting to a non--conducting state. 
We furthermore show that the vanishing of the current antiresonance due to
dephasing of the coupled groundstates coherence (which can be controlled by tuning the tunnel coupling)
can be used to obtain quantitative estimates for inelastic dephasing rates in coupled dots.

\begin{figure}[ht]
\unitlength1cm
\begin{picture}(9,7)
\epsfxsize=9cm
\put(-0.5,0.5){\epsfbox{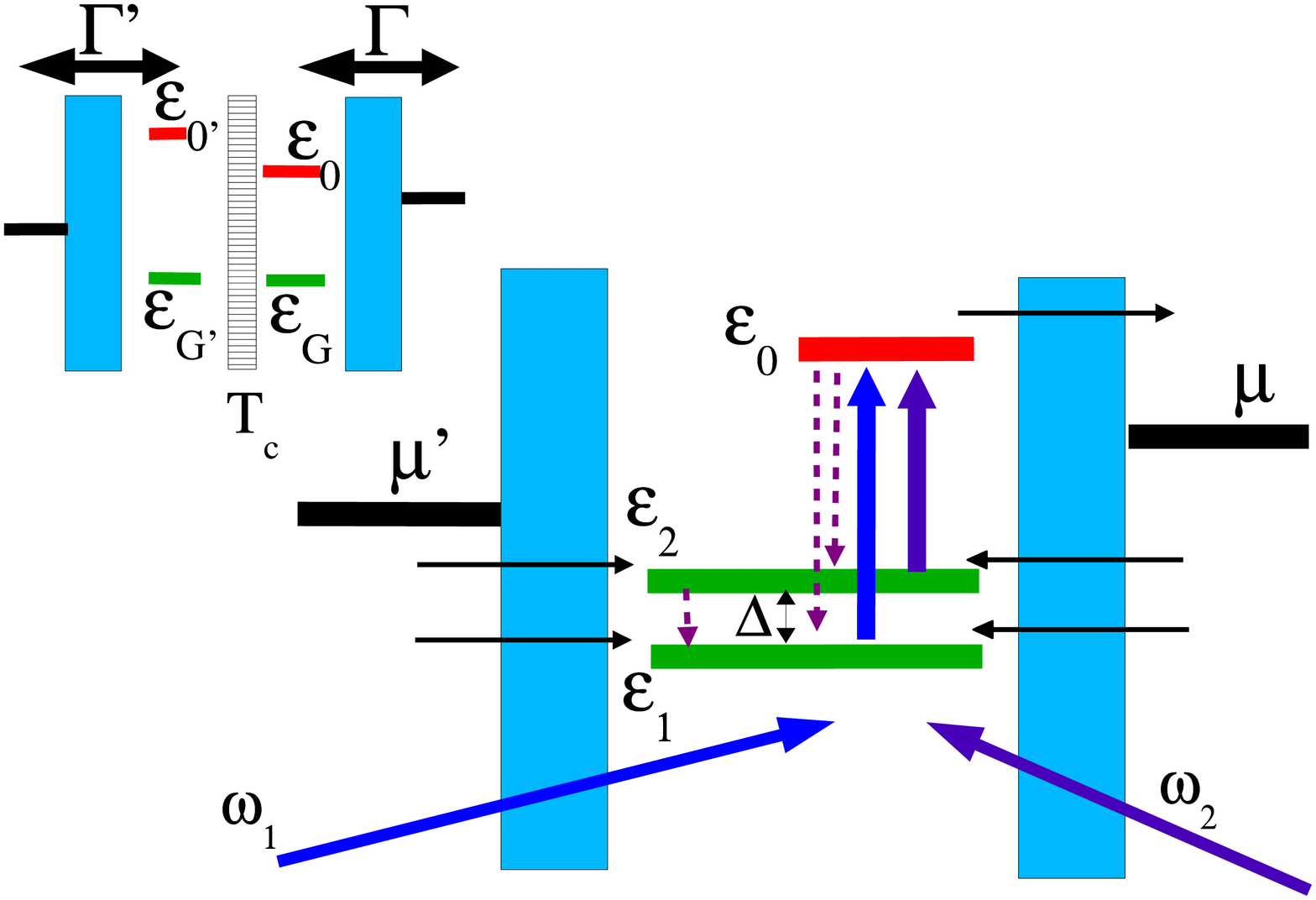}}
\end{picture}
\caption[]{\label{ddot.eps}Level scheme for two coupled quantum dots in Coulomb blockade regime.
Two tunnel coupled groundstates $|G\rangle$ and $|G'\rangle$ (small inset) form states
$|1\rangle$ and $|2\rangle$ from which an electron
is pumped to the excited state $|0\rangle$ by two light sources of frequency $\omega_1$ and $\omega_2$.
Relaxation by acoustic phonon emission is indicated by dashed arrows.}
\end{figure}

The effect appears in double quantum dots 
where electron transport involves tunneling through two bonding and antibonding ground states $|1\rangle$
and $|2\rangle$ and one additional excited state $|0\rangle$, see Fig. \ref{ddot.eps}. 
Leads coupled to both dots have chemical potentials such that
electrons can tunnel into the groundstates but leave the dot only
through the excited state. The system is driven by two light (microwave) sources 
with frequencies $\omega_1$ and $\omega_2$ that are detuned
off the two excitation energies by $\hbar\delta_1:=\varepsilon_0-\varepsilon_1-\hbar\omega_1$
and $\hbar\delta_2:=\varepsilon_0-\varepsilon_2-\hbar\omega_2$. Relaxation from the excited
level by acoustic phonon emission traps the dot in a coherent superposition of 
the bonding and the antibonding state, if $\delta_R:=\delta_2-\delta_1$ is tuned to zero. 
In this case, the excited level becomes completely depopulated. 
In the case  of real atoms, 
the resulting trapping of the electron in a radiatively decoupled coherent superposition  
leads to `dark resonances' in the fluorescence emission.
In the double dot case discussed here, the dark resonance effect appears 
as a suddenly vanishing electron current for $\delta_R=0$. 
We suggest that for low enough microwave
intensity, the effect can serve as a very sensitive, optically controlled current switch.

Atomic dark states have been found to be extraordinary stable against a number of perturbations \cite{delignie}.
In the quantum dot case, due to the Pauli blocking of the leads, a trapped electron can not tunnel out of 
the ground state coherent superposition. Furthermore, this superposition
is protected from incoming electrons due to Coulomb blockade (no second electron can tunnel in).
These two mechanisms guarantee the robustness of the effect, which is limited
only by dephasing from inelastic processes. The latter are due
to spontaneous emission of  phonons in double dots \cite{FujetalTaretal,BK99}
and can be controlled by tuning system parameters with gate voltages.

In our model, we consider a double quantum dot 
in the strong Coulomb blockade regime that is determined 
by transitions between states of fixed particle number $N$ and $N+1$.
The two tunnel coupled $N+1$--particle
groundstates $|G\rangle$ and $|G'\rangle$ (see Fig. \ref{ddot.eps}, inset)
have energy difference $\varepsilon:=\varepsilon_G-\varepsilon_{G'}$
and hybridize into states $|1\rangle$ and $|2\rangle$ of energy
difference $\Delta:=\varepsilon_2-\varepsilon_1=(\varepsilon^2+4T_c^2)^{1/2}$. Here, $T_c$ denotes the
tunnel coupling matrix element. 
The system is irradiated with two coherent microwave sources
with frequencies $\omega_1$ and $\omega_2$, driving the
transitions $|1\rangle \to |0\rangle$ and $|2\rangle \to |0\rangle$.
Here, $|0\rangle$  is the first excited state of the same electron
number $N+1$ in the right dot with energy $\varepsilon_0$.
Furthermore, the energy of the first excited level $|0'\rangle$ of the other (left) dot is assumed to be 
in off--resonance for transitions to and from the two ground states. If the energy difference
$\varepsilon_{0'}-\varepsilon_0$ is much larger than $T_c$, the hybridization of $|0'\rangle$ 
with $|0\rangle$ can be neglected. 

The microwave radiation pumps electrons into the excited level $|0\rangle$ such that
transport through $N+1$ particle states becomes possible if both dots are connected to reservoirs
of free two--dimensional electrons.
We assume the Coulomb charging energy $U$ to be so large that
states with two additional electrons can be neglected.
Typical values are  1meV $\lesssim U \lesssim$ 4meV in lateral double dots \cite{FujetalTaretal}.
The chemical potentials $\mu$ and $\mu'$ are tuned slightly above
$\varepsilon_2$; this excludes the co--tunneling like reentrant resonant 
tunneling process that can exist in three--level dots \cite{Kuzetal96}.

The light coupling is described by an interaction Hamiltonian in the 
rotating wave approximation,
\begin{equation}\label{H_p}
H_I(t)=-\frac{\hbar \Omega_1}{2}e^{-i\omega_1 t}|0\rangle\langle 1| -\frac{\hbar
\Omega_2}{2}e^{-i\omega_2 t}|0\rangle\langle 2| + h.c.,
\end{equation}
where non--resonant terms have been neglected and 
$\Omega_j=({E}_j/{\hbar})\langle 0|ez|j\rangle$, $j=1,2$,
are the Rabi frequencies,
where $E_j$ is the projection of the electric field vectors of the
light onto the dipole moments for the transitions $1\to 0$, $2\to 0$. 
The coupling of the dot groundstates to the leads is described by the standard tunnel Hamiltonian
\begin{equation}
H_V=\sum_{{\bf k}i=G,G'}\left(V_{{\bf k}i}c_{{\bf k}i}^{\dagger}|E\rangle \langle i| 
+ c.c.\right)
\end{equation}
and correspondingly for the excited state $|0\rangle$.
Here, $|E\rangle$ denotes the `empty' double dot $N$--particle state before tunneling of an additional
electron, $c_{{\bf k}i}^{\dagger}$ creates
an electron with quantum number ${\bf k}$ in the reservoir connected to the dot groundstate
$i=G$ or $i=G'$, and $V_{{\bf k}i}$  denotes the corresponding tunnel matrix element. 
The rates $\Gamma$ (right dot) and $\Gamma'$ (left dot) for tunneling between the dots and the
connected reservoirs can be calculated from $H_V$
by second order perturbation theory. If the chemical potentials $\mu$ and $\mu'$ 
are as indicated in Fig.(\ref{ddot.eps}), electron tunneling occurs 
by {\em in}--tunneling that changes $|E\rangle$ into $|G\rangle$ 
at a rate $\Gamma$ and $|E\rangle$ into
$|G'\rangle$ at the rate $\Gamma'$,
whereas {\em out}--tunneling from $|G\rangle$ and $|G'\rangle$  is Pauli blocked. 
The corresponding rates $\gamma_1$ and $\gamma_2$ for tunneling into the
hybridized states $|1\rangle$ and $|2\rangle$ are
$\gamma_{1,2}=[(\Delta \pm \varepsilon )^2 \Gamma +4T_c^2 \Gamma']/ [(\Delta \pm \varepsilon )^2 +4T_c^2 ]$.
On the other hand, electrons can leave the dot only
by tunneling {\em out of} the state $|0\rangle$ (but not in) at the rate $\Gamma$.
This tunneling is only into the right lead because we assumed negligeable hybridization of $|0\rangle$ with
$|0'\rangle$.
Here and in the following, we neglect the energy dependence of $\Gamma$ and $\Gamma'$ for simplicity. 

In coupled quantum dots, decay of excited levels is due to spontaneous emission
of phonons rather than photons \cite{FujetalTaretal}. We denote the corresponding 
decay rates for the state $|0\rangle$ and $|2\rangle$ by $\Gamma^0$ and $\Gamma_{21}$,
respectively. The lowest state $|1\rangle$ is stable against decay.
For the moment, we take these rates as given and discuss quantitative estimates below. 
We are then in the position to set up equations of motion for the time-dependent 
occupation probabilities $p_j(t)$, $j=E,1,2,0$, of the four double dot states.
The spontaneous phonon emission and the single electron tunneling gives rise to an
incoherent dynamics, while the electron--light interaction in treated fully
coherently. One has
\begin{eqnarray}
\label{eq:peqns}
\dot{p}_E&=&-(\gamma_1+\gamma_2) p_E + {\Gamma} p_0\nonumber\\
\dot{p}_0&=&-(\Gamma^0+\Gamma) p_0
+\Im{\rm m}\left(\Omega_1\tilde{\rho}_{10}+\Omega_2\tilde{\rho}_{20}\right)\nonumber\\
\dot{p}_1&=&\alpha_1\Gamma^0 p_0+\gamma_1 p_E +\Gamma_{21}p_2 
-\Im{\rm m}\left(\Omega_1\tilde{\rho}_{10}\right)\nonumber\\
\dot{p}_2&=&\alpha_2\Gamma^0 p_0+\gamma_2 p_E -\Gamma_{21}p_2 
-\Im{\rm m}\left(\Omega_2\tilde{\rho}_{20}\right).
\end{eqnarray}
Here, $\alpha_1=1-\alpha_2=(\Delta+\varepsilon)^2/[(\Delta+\varepsilon)^2+4T_c^2]$ and
$\tilde{\rho}_{0j}=\tilde{\rho}_{j0}^*=\rho_{0j}e^{i\omega_j t}$ are slowly-varying off--diagonal
matrix elements of the reduced density operator of the double dot, whose
equations of motion close the set (\ref{eq:peqns}). One has
\begin{eqnarray}\label{offequation}
\dot{\tilde{\rho}}_{10}&=&
-D_1\tilde{\rho}_{10}+i\frac{\Omega_1^*}{2}\left(p_1-p_0\right)
+i\frac{\Omega_2^*}{2}\tilde{\rho}_{12}\nonumber\\
\dot{\tilde{\rho}}_{02}&=&
-D_2\tilde{\rho}_{02}-i\frac{\Omega_2}{2}\left(p_2-p_0\right)
-i\frac{\Omega_1}{2}\tilde{\rho}_{12}\nonumber\\
\dot{\tilde{\rho}}_{12}&=&
-[i\delta_R+\Gamma_{21}/2]\tilde{\rho}_{12}-i\frac{\Omega_1^*}{2}\tilde{\rho}_{02}
+i\frac{\Omega_2}{2}\tilde{\rho}_{10}, 
\end{eqnarray}
where we defined resonance denominators 
$D_j:=(-1)^j i\delta_j+ \alpha_j \Gamma^0+ \Gamma/2$ 
that appear in the solution for the coherences in the stationary case for large
times which we consider from now on. 
Together with the normalization condition $p_E+p_1+p_2+p_0=1$, the
stationary solution is then easily obtained.

Before discussing the stationary tunnel current, we estimate the inelastic rates
$\Gamma^0$ and $\Gamma_{21}$ which determine if or if not the effect can be
observed in quantum dots at all. In the following, we restrict ourselves to
lateral dots. Relaxation from the excited dot level $|0\rangle$
is due to acoustic phonon emission at a rate 
\begin{equation}
\Gamma^0=(2\pi/\hbar)\sum_{\bf Q}
|\lambda_{\bf Q}|^2 \delta (\hbar\omega_{Q}-\varepsilon_0)F_z(q_z)G(q_{\|}),
\end{equation} 
where $\lambda_{\bf Q}$ is the deformation potential matrix element, ${\bf Q}=({\bf q}_{\|},q_z)$ the phonon
wave vector, $\omega_Q=c|{\bf Q}|$, and $F_z$ and $G$ are the quantum well and lateral dot form factor
which cut off phonons with $|q_z|\gtrsim l_z^{-1}$ and $|{\bf q}_{\|}|\gtrsim l^{-1}$, where
$l_z$ is the quantum well width and $l$ an estimate for the dot diameter.
For $\varepsilon_0 \lesssim 0.5$meV and a typical well width of $l_z\sim 50 \AA$, only the
lateral cutoff $G$ is effective here at energies above $\hbar \omega_l=\hbar c/l$, where $c$ is the
longitudinal speed of sound \cite{BB90}.
The explicit form of $G$ depends on the shape of the many--electron
wave functions $\langle {\bf x}|0\rangle$ and $\langle {\bf x}|i\rangle$, $i=1,2$ and is never
known exactly for realistic dots with $N\gtrsim 10$ electrons. Assuming a 
form $G(q)=(ql)^2/[1+(ql)^2]^2$ that smoothly interpolates between $G(0)=G(\infty)=0$ and using
material parameters for GaAs and $l = 200$ nm, we find rates between $\Gamma^0(\varepsilon_0=0.5 $meV$)=
6 \cdot 10^8$ s$^{-1}$ and $\Gamma^0(\varepsilon_0=10 $meV $)= 2 \cdot 10^{10}$ s$^{-1}$. 
\begin{figure}[t]
\unitlength1cm
\begin{picture}(9,7)
\epsfxsize=9cm
\put(-0.5,0.5){\epsfbox{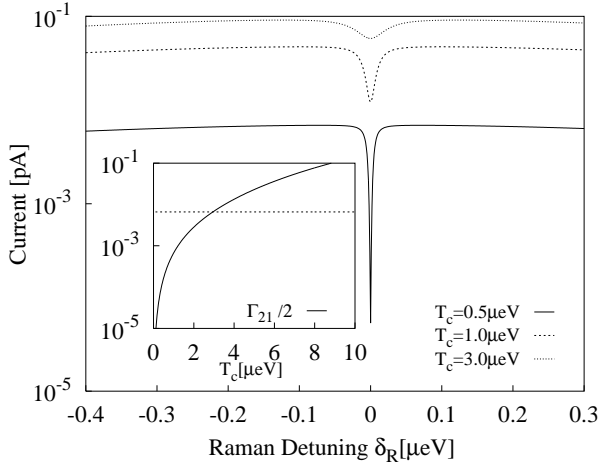}}
\end{picture}
\caption[]{\label{cu_eps0.eps}Tunnel current antiresonance through 
double dot system from Fig.\ref{ddot.eps} with groundstate energy difference $\varepsilon=10\mu$eV.
The Rabi frequencies $\Omega_1$ and $\Omega_2$ are taken to be equal,
parameters are $\Omega_R=0.2\Gamma^0$,
$\Gamma=\Gamma'=\Gamma^0=10^9$s$^{-1}$, 
where $\Gamma^0$ is the relaxation rate due to acoustic phonon emission from $|0\rangle$.
Inset: Inelastic rate $\Gamma_{21}$ (in $\mu eV/\hbar$), Eq.(\ref{Gamma21}), with $\hbar \omega_d=20 \mu$eV.
Dashed line indicates the crossover at $\Gamma_{21}/2=|\Omega_R|^2/2[\Gamma^0+\Gamma]$, cf. 
Eq.(\ref{delta12}).}
\end{figure} 
Most important for the observation of the population trapping effect
in dots is the relaxation rate $\Gamma_{21}$. 
In GaAs/AlGaAs lateral double dots,
$\Gamma_{21}$ is mainly due to the spontaneous emission
of phonons \cite{FujetalTaretal,BK99}. 
In experiments, gate voltages can be applied to tune the ground state level splitting to small values. 
Here, we assume $\Delta\lesssim 20 \mu$eV where form factor cutoffs are no longer effective.
One obtains
\begin{equation}\label{Gamma21}
\Gamma_{21}(\Delta)\approx 2\pi \left(\frac{T_c}{\Delta}\right)^2 g \frac{\Delta}{\hbar}
\left[1-\frac{\sin (\Delta/\hbar \omega_d)}{\Delta/\hbar \omega_d}\right],
\end{equation}
where $\omega_d:=c/d$, $g \lesssim 0.05$ the dimensionless coupling constant,
$d$ is the distance between the dot centers,
and we assumed identical shapes of both dots for simplicity and neglected
the small overlap between the states $|G\rangle$ and $|G'\rangle$.
Furthermore, a simplified model with bulk piezoelectric phonons has been adopted. 
Important here is that in contrast to real atoms the 
spontaneous rate $\Gamma_{21}$ can be {\em tuned} in gated double dots by varying $T_c$ and/or
$\varepsilon$. 
This allows one to study how the coherent superposition of states
is destroyed due to the interaction with the phonon bath as discussed now.
 
\begin{figure}[ht]
\unitlength1cm
\begin{picture}(9,7)
\epsfxsize=9cm
\put(-0.5,0.5){\epsfbox{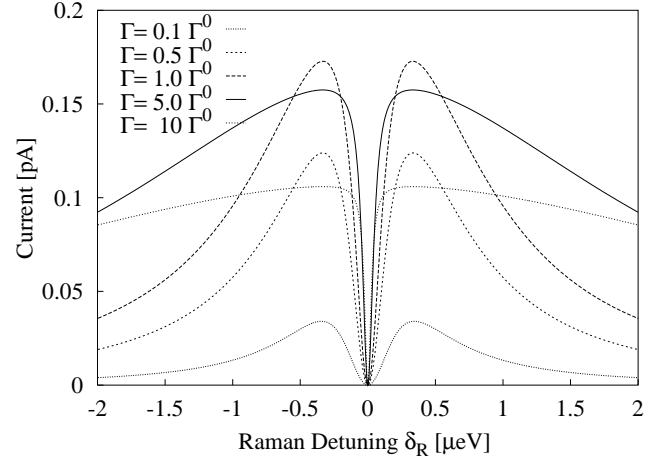}}
\end{picture}
\caption[]{\label{cu1.eps}Current for fixed coupling $T_c$ and 
different tunnel rates $\Gamma=\Gamma'$. Parameters are $\varepsilon=10\mu$eV,
$\Gamma^0=10^9$s$^{-1}$, $\Omega_R=1.0\Gamma^0$, $T_c=1\mu$eV.}
\end{figure} 
The stationary electric current $I$
is obtained from the net flow of electrons with charge $-e<0$ through
either of the tunnel barriers connecting the dot to the reservoirs, 
$I=-e\Gamma  [p_0-p_E]_{\rm stat}=-e\Gamma'[ p_E]_{\rm stat}$. 
Fig. \ref{cu_eps0.eps} shows the result for $I$ as a function
of the Raman detuning $\delta_R$ for $\Omega_1=\Omega_2$ and ground state
energy difference $\varepsilon=10\mu$eV. 
Our calculations have been done for zero temperature $T=0$. For finite $T$, reabsorption of phonons
which would smear the ground state levels can be suppressed by choosing a sufficiently large
$\varepsilon \gtrsim k_B T$. 

Close to $\delta_R=0$, the overall Lorentzian profile 
breaks in and shows a sharp current antiresonance. For fixed microwave intensity (fixed Rabi
frequency $\Omega_R:=(\Omega_1^2+\Omega_2^2)^{1/2}$)
and increasing tunnel coupling $T_c$,
the inelastic rate $\Gamma_{21}$, Eq.(\ref{Gamma21}), increases (inset). As a result,
the antiresonance becomes broader and finally disappears for larger tunnel coupling $T_c$. 
The half--width $\delta_{1/2}$ of the current antiresonance can be found via
the stationary solution of Eq.(\ref{eq:peqns}) and Eq.(\ref{offequation}) 
from the pole of a two--photon denominator as a function of $\delta_R$. 
We find for the symmetric case $\varepsilon=0$
\begin{equation}\label{delta12}
\delta_{1/2}\approx\frac{\Gamma_{21}}{2}+\frac{|\Omega_R|^2}{2[\Gamma^0+\Gamma]}.
\end{equation}
Thus, $\delta_{1/2}$ increases with the inelastic rate $\Gamma_{21}$.
For fixed microwave intensity, the vanishing of the antiresonance 
sets in for $\Gamma_{21} \gtrsim |\Omega_R|^2/[\Gamma^0+\Gamma]$, cf. the inset of Fig. \ref{cu_eps0.eps}.
On the other hand, with {\em increasing} elastic tunneling $\Gamma$ out of the dot
we reckognize the striking fact that $\delta_{1/2}$ {\em decreases} down to its lower limit $\Gamma_{21}/2$.
This behavior is shown in Fig. \ref{cu1.eps}. For increasing tunnel rate $\Gamma$, the
current increases until an overall maximal value is reached at $\Gamma\approx \Gamma^0$.
The curve $I(\delta_R)$ decreases again and becomes very broad
if the elastic tunneling becomes much faster than the inelastic relaxation $\Gamma^0$.
Simultaneously, the center antiresonance then becomes sharper and sharper with increasing
$\Gamma$, its halfwidth $\delta_{1/2}$ approaching the limit $\Gamma_{21}/2$, Eq. (\ref{delta12}).

The appearance of the sharp current antiresonance is due to a trapping
of the additional electron in a coherent superposition of the two ground states
$|1\rangle$ and $|2\rangle$ that decouples from the light. One can define 
linear combinations \cite{Arimondo96} $|NC(t)\rangle:=(\Omega_2/\Omega_R)|1\rangle
 - (\Omega_1/\Omega_R)e^{i(\omega_2-\omega_1)t}|2\rangle$ and the orthogonal state
$|C(t)\rangle$.
At Raman resonance, only $|C(t)\rangle$ couples to the light, and
excitation of the electron from $|C(t)\rangle$ to the excited state $|0\rangle$
with a subsequent decay into $|C(t)\rangle$ {and} $|NC(t)\rangle$ gradually pumps
all the population into $|NC(t)\rangle$. This is  because in the latter state the 
electron is decoupled from the light and can not be excited again. 

We point out that the resonance effect described here differs physically from 
other transport effects in AC--driven systems, such as coherent
destruction of tunneling \cite{GrossmannWagner}, tunneling through photo--sidebands
\cite{BruderSchoellerInarrea}, or coherent pumping of electrons \cite{SW96,WS99}. These 
phase--coherent effects are due to an additional time--dependent phase 
that electrons pick up while tunneling. Then, the time evolution within the system is ideally
completely coherent with dissipation being a disturbance rather then necessary for the effect to occur.
In contrast, the trapping effect discussed here requires incoherent relaxation
(phonon emission) within the system in order to create the trapped coherent superposition of the
ground states. 

To conclude, our results suggest that the population trapping effect can be observed in the 
tunnel current through double dots irradiated with two microwaves. 
It offers the possibility to switch a current optically and to determine 
the interdot inelastic rate $\Gamma_{21}$ from the antiresonance linewidth $\delta_{1/2}$, Eq.(\ref{delta12}). 
The microwave frequencies $\nu$ should be
such that the first excited level in one of the dots is coupled by one--photon processes to the groundstates.
An estimate with a single particle excitation energy of $\delta\varepsilon = 0.5$meV yields
$\nu = \delta\varepsilon /h =$120 GHz which should be attainable with present day technology. 
The Raman shift $\delta_R\equiv\delta_2-\delta_1$ can  be scanned through by
fixing one of the frequencies (e.g., $\omega_1$) at resonance such that $\delta_1\equiv0$,
and changing $\omega_2$ and therewith
$\delta_R=\omega_1-\omega_2-\Delta/\hbar$.
Both the relaxation rate $\Gamma^0$ and the dephasing rate $\Gamma_{21}$ then can be obtained 
from  $I(\delta_R)$--curves for different values of, e.g., the tunnel coupling $T_c$ or the 
energy difference $\varepsilon$. 

Finally, we comment on
the dephasing channel due to tunneling of electrons from the ground state
coherent superposition into holes created by absorption of photons in the leads.
The rate $\Gamma_r$ for such processes is proportional to $(\Omega_R/2\pi\nu)^2$  \cite{Bra97} and turns out
to be at least one order of magnitude less than the intrinsic dephasing rate
$\Gamma_{21}$ unless one tunes to very small  tunnel couplings $T_c \lesssim 0.5\mu$eV.
In this regime,
$\Gamma_r$ starts to dominate over $\Gamma_{21}$, and the halfwidth $\delta_{1/2}$ then becomes
independent of $T_c$.

This work has been supported by  the TMR network
`Quantum transport in the frequency and time domains',
DFG projects Kr 627/9--1, Br 1528/4--1 (T. B.), and DFG project Li 417/4--1 (F. R.).


\begin{thebibliography}{10}




\bibitem{Bonetal98}
{N.~H.~Bonadeo, J.~Erland, D.~Gammon, D.~Park, D.~S.~Katzer, and D.~G.~Steel},
  Science {\bf 282},  1473  (1998).

\bibitem{Blietal98a}
{R.~H.~Blick, D.~W.~van~der~Weide, R.~J.~Haug, and K.~Eberl}, Phys. Rev. Lett.
  {\bf 81},  689  (1998).

\bibitem{Kouetal94}
{L. P. Kouwenhoven, S. Jauhar, K. McCormick, P. L. McEuen, Yu. V. Nazarov, N.
  C. van der Vaart, and C. T. Foxon}, Phys. Rev. B {\bf 50},  2019  (1994).

\bibitem{Oosetal98}
{T. H. Oosterkamp, T. Fujisawa, W. G. van der Wiel, K. Ishibashi, R. V. Hijman,
  S. Tarucha, and L. P. Kouwenhoven}, Nature {\bf 395},  873  (1998).

\bibitem{Arimondo96}
{E. Arimondo},  in {\em Progress in Optics}, edited by E. Wolf (Elsevier,
  Braunschweig, 1996), Vol.~35, p.\ 257.

\bibitem{delignie}
{ M.C. de Lignie and E.R. Eliel, Opt. Comm. {\bf 72}, 205 (1989); E.R. Eliel,
  Adv. At. Mol Phys. {\bf 30}, 199 (1993)}.

\bibitem{FujetalTaretal}
{T.~Fujisawa, T.~H.~Oosterkamp, W.~G.~van~der~Wiel, B.~W.~Broer, R.~Aguado,
  S.~Tarucha, and L.~P.~Kouwenhoven, Science {\bf 282}, 932 (1998); S.~Tarucha,
  T.~Fujisawa, K.~Ono, D.~G.~Austin, T.~H.~Oosterkamp, and W.~G.~van der Wiel,
  Microelectr. Engineer. {\bf 47}, 101 (1999).}

\bibitem{BK99}
T. Brandes and B. Kramer, Phys. Rev. Lett. {\bf 83},  3021  (1999).

\bibitem{Kuzetal96} 
V. V. Kuznetsov, A. K. Savchenko, M. E. Raikh, L. I. Glazman,
D. R. Mace, E. H. Linfield, D. A. Ritchie,
Phys. Rev. B {\bf 54}, 1502 (1996).


\bibitem{BB90}
U. Bockelmann and G. Bastard, Phys. Rev. B {\bf 42},  8947  (1990).

\bibitem{GrossmannWagner}
{F. Grossmann, T. Dittrich, P. Jung, and P. H\"anggi, Phys. Rev. Lett. {\bf
  67}, 516; M. Wagner, Phys. Rev. {\bf A 51}, 798 (1995).}

\bibitem{BruderSchoellerInarrea}
{C.~Bruder and H.~Schoeller, Phys. Rev. Lett. {\bf 72}, 1076 (1994); J.
  I\~narrea, G. Platero, and C. Tejedor, Phys. Rev. {\bf B 50}, 4581 (1994);
  Ph. Brune, C. Bruder, and H. Schoeller, Phys. Rev. {\bf B 56}, 4730 (1997)}.

\bibitem{SW96}
C.~A. Stafford and N.~S. Wingreen, Phys. Rev. Lett. {\bf 76},  1916  (1996).

\bibitem{WS99}
{M. Wagner and F. Sols}, Phys. Rev. Lett. {\bf 83},  4377  (1999).


\bibitem{Bra97}
T. Brandes, Phys. Rev. {\bf B 56}, 1213 (1997).

\end{thebibliography}
\end{document}